\documentclass[preprint2]{aastex63}
\usepackage{fontawesome}
\usepackage{CJK}
\received{\today}
\revised{NA}
\accepted{NA}
\submitjournal{AAS Journal}

\shorttitle{Dual-Aperture FN}
\shortauthors{Wang et al.}
\graphicspath{{./}{figures/}}

\begin{document}
\begin{CJK*}{UTF8}{gbsn}

\title{Exoplanets Sciences with Nulling Interferometers \\and a Single-Mode Fiber-Fed Spectrograph}

\correspondingauthor{Ji Wang}
\email{wang.12220@osu.edu}

\author[0000-0002-4361-8885]{Ji Wang (王吉)}
\affiliation{Department of Astronomy, The Ohio State University, 100 W 18th Ave, Columbus, OH 43210 USA}


\author[0000-0002-5018-7761]{Colby Jurgenson}
\affiliation{Department of Astronomy, The Ohio State University, 100 W 18th Ave, Columbus, OH 43210 USA}










\begin{abstract}

Understanding the atmospheres of exoplanets is a milestone to decipher their formation history and potential habitability. High-contrast imaging and spectroscopy of exoplanets is the major pathway towards the goal. Directly imaging of an exoplanet requires high spatial resolution. Interferometry has proven to be an effective way of improving spatial resolution. However, means of combining interferometry, high-contrast imaging, and high-resolution spectroscopy have been rarely explored. To fill in the gap, we present the dual-aperture fiber nuller (FN) for current-generation 8-10 meter telescopes, which provides the necessary spatial and spectral resolution to (1) conduct follow-up spectroscopy of known exoplanets; and (2) detect planets in debris-disk systems. The concept of feeding a FN to a high-resolution spectrograph can also be used for future space and ground-based missions. We present a case study of using the dual-aperture FN to search for biosignatures in rocky planets around M stars for a future space interferometry mission. Moreover, we discuss how a FN can be equipped on future extremely large telescopes by using the Giant Magellan Telescope (GMT) as an example. 

\end{abstract}



\section{Introduction} \label{sec:intro}

Direct imaging and spectroscopy of exoplanets provides a wealth of data sets to understand planet orbital dynamics and atmospheric compositions. Current-generation instruments can detect planets that are $\sim10^6$ times fainter than the host star at sub-arcsec separation~\citep{Macintosh2015, Keppler2018}. In parallel, direct spectroscopy of sub-stellar companions with high-resolution spectrographs (R$>$20,000) becomes an emerging field, which opens the window for probing atmospheric circulation~\citep{Snellen2010}, surface inhomogeneity~\citep{Crossfield2014}, and planet rotation~\citep{Schwarz2016, Bryan2018}. 

Combining high-contrast imaging and high-resolution spectroscopy is logically the next step to improve sensitivity and broaden the science scope of direct imaging and spectroscopy. We use the term high dispersion coronagraphy (HDC) for the combination of the two techniques~\citep{Wang2017, Mawet2017}. 

HDC invokes multiple stages to suppress stellar light and extract the planet's signal. Specifically, high-contrast imaging suppresses stellar light and spatially separates the planet from its host star. A single-mode fiber injection system filters out stellar noise at the planet location since the electric field of a stellar speckles does not couple to the fundamental mode of a single-mode fiber. High-resolution spectroscopy further distinguishes the planet signal from stellar signal by its unique spectral features such as absorption lines and radial velocity. Using this three-pronged starlight suppression, HDC can achieve the high sensitivity  to study terrestrial planets in the habitable zone~\citep{Kawahara2014b,Lovis2017, Wang2017, Mawet2017, Wang2018}. 

Alternatively, an interferometer can be used to effectively suppress starlight. In contrast to a co-axial beam combiner, which is used for the Keck Interferometer~\citep{MillanGabet2011, Mennesson2014} and the Large Binocular Telescope Interferometer~\citep{Ertel2018, Ertel2020}, a multi-axial beam combiner maximizes the spatial resolution, e.g., the Palomar fiber nuller~\citep[PFN, ][]{Haguenauer06, Mennesson2011,Serabyn2019} and the Fizeau imaging mode at LBT~\citep{Spalding2018}. We will focus on the multi-axial interferometry because of its enhanced spatial resolution and its potential of feeding a high resolution spectrograph with a single-mode fiber. 

A more recent development of HDC is the vortex fiber nuller~\citep[VFN, ][]{Ruane2018}. VFN provides a unique solution for coronagraphy and high-resolution spectroscopy at sub $\lambda$/D angular resolution for next generation ground-based extremely large telescopes~\citep[ELTs, ][]{Ruane2019}, and the concept has been demonstrated~\citep{Echeverri2019}.

While ELTs with full capability of HDC are a decade away, we present in this paper a FN concept that can be applied to current-generation 8-10 meter telescopes, namely the dual-aperture FN. The concept---combining interferometry with high-resolution spectroscopy---has the potential to expedite the science goal of direct spectroscopy of exoplanets at tens of mas separations.

The dual-aperture FN can also be a choice for future space missions in search for habitable planets and biosignatures in their atmospheres, especially for planets around M stars. Spatial resolution is the major limiting factor that prevents space missions from pursuing direct spectroscopy of habitable planets around M stars. The dual-aperture FN permits (1) sufficient angular resolution with long-base line interferometry; and (2) searching for biosignatures in near infrared where their spectral features are abundant.  

We will introduce a dual-aperture FN in \S \ref{sec:sim} and evaluate its performance in \S \ref{sec:metrics}. Science cases that are enabled by the dual-aperture FN are discussed in \S \ref{sec:sci}. A comparison between FN and VFN is given in \S \ref{sec:fn_vfn}. Our findings are summarized in \S \ref{sec:summary}.




\begin{figure*}[t!]
\plotone{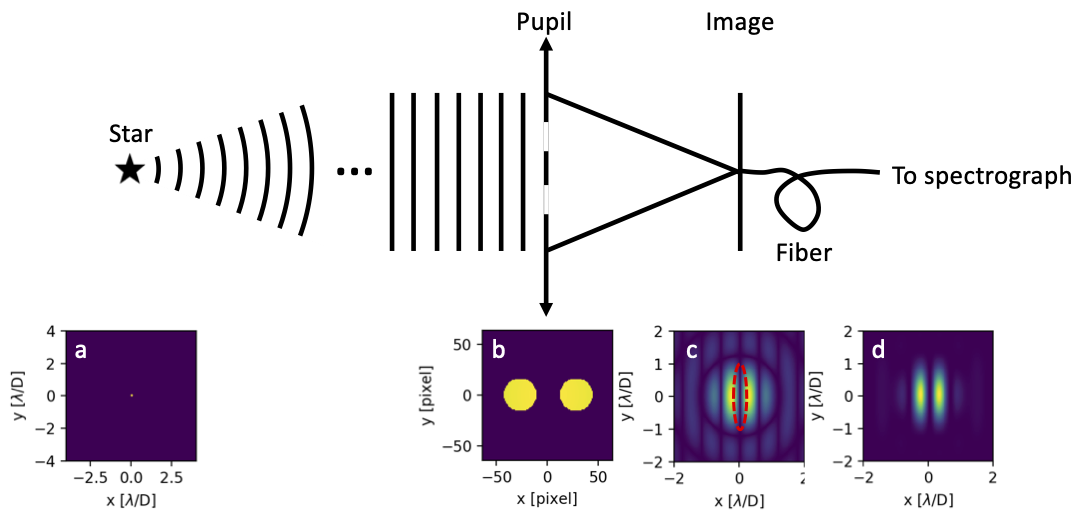}
\caption{{Illustration of a dual-aperture FN. {\bf{(a)}}: Intensity of a point source that is located at an infinite distance, i.e., an unresolved star. {\bf{(b)}}: A dual-aperture pupil plane with a $\pi$ phase offset. {\bf{(c)}}: An image of the interferogram of the dual-aperture. Red dashed circle is the ``footprint" of an on-axis single-mode fiber core after a beam-shaping device. {\bf{(d)}}: Coupling efficiency map vs. angular separations from the optical axis (i.e., coordinate [0, 0]).
}
\label{fig:vfn_dual}}
\end{figure*}

\section{Simulation}
\label{sec:sim}

\subsection{A Dual-Aperture FN}

The dual-aperture FN serves as a bridge between current 8-10 meter telescopes and future ELTs. For example, Keck telescopes, the Large Binocular Telescope, and the Very Large Telescopes Interferometer (VLTI) are all capable of dual-aperture interferometry, providing spatial resolutions that are comparable or even better than those from ELTs. However, ELTs offer superior photon collecting power to existing facilities. 

Our dual-aperture FN concept is illustrated in Fig. \ref{fig:vfn_dual}. We use a dual-aperture that is similar to the large binocular telescope interferometer~\citep[LBTI, ][]{Hinz2016}, with a baseline-aperture ratio of 22.8 m / 8.4 m = 2.71. The concept can be generalized to any dual-aperture interferometers such as Keck and VLTI.

\subsection{Coupling Efficiency}

{The coupling efficiency is the overlapping integral of the EM field and the mode profile of a single-mode fiber~\citep{Wagner1982,Jovanovic2017}:
\begin{equation}
\label{eq:eta}
    \eta=\frac{\left|\int{E(r, \theta)\Psi(r, \theta)dA}\right|^2}{\int{\left|E(r, \theta)\right|^2dA}\cdot\int{\left|\Psi(r, \theta)\right|^2dA}},
\end{equation}
where $\eta$ is the fiber coupling efficiency, $E(r, \theta)$ is the electromagnetic (EM) field and $\Psi(r, \theta)$ is the mode profile of a single-mode fiber.

The point spread function (PSF) of the dual-aperture is highly non-gaussian (shown in Panel (c) in Fig. \ref{fig:vfn_dual}), a beam shaping device is therefore needed to improve the coupling efficiency into a single-mode fiber. To do so, the PSF and the fiber mode profile need to be matched. We assume an optical device (e.g., a pair of cylindrical lenses or an aspheric lens) to change the aspect ratio of the PSF to $\sim$1:1. 

This is equivalent to feed the EM field to an elongated two-dimensional gaussian beam (shown as the red dashed circle in panel (c) in Fig. \ref{fig:vfn_dual}), which is the ``footprint" of the fundamental mode of the on-axis single-mode fiber after the beam-shaping device. The aspect ratio of the gaussian beam is determined by the baseline-aperture ratio, for which we adopt the LBTI value of 2.71.

The coupling efficiency peaks at 35.3\% at 0.28 $\lambda/D$ or 0.76 $\lambda/B$, where $\lambda$ is wavelength, $D$ is the sub-aperture size, and $B$ is the edge-to-edge baseline. Coupling efficiency as a function of angular separation is shown in Fig. \ref{fig:eta_p_sep}. The region with at least half of the peak efficiency goes from 0.36 to 1.12 $\lambda/B$. The maximum throughput as a function of the gaussian $\sigma$ (along the elongation direction) is shown in Fig. \ref{fig:eta_p_sep}.     

\begin{figure*}[t!]
\plotone{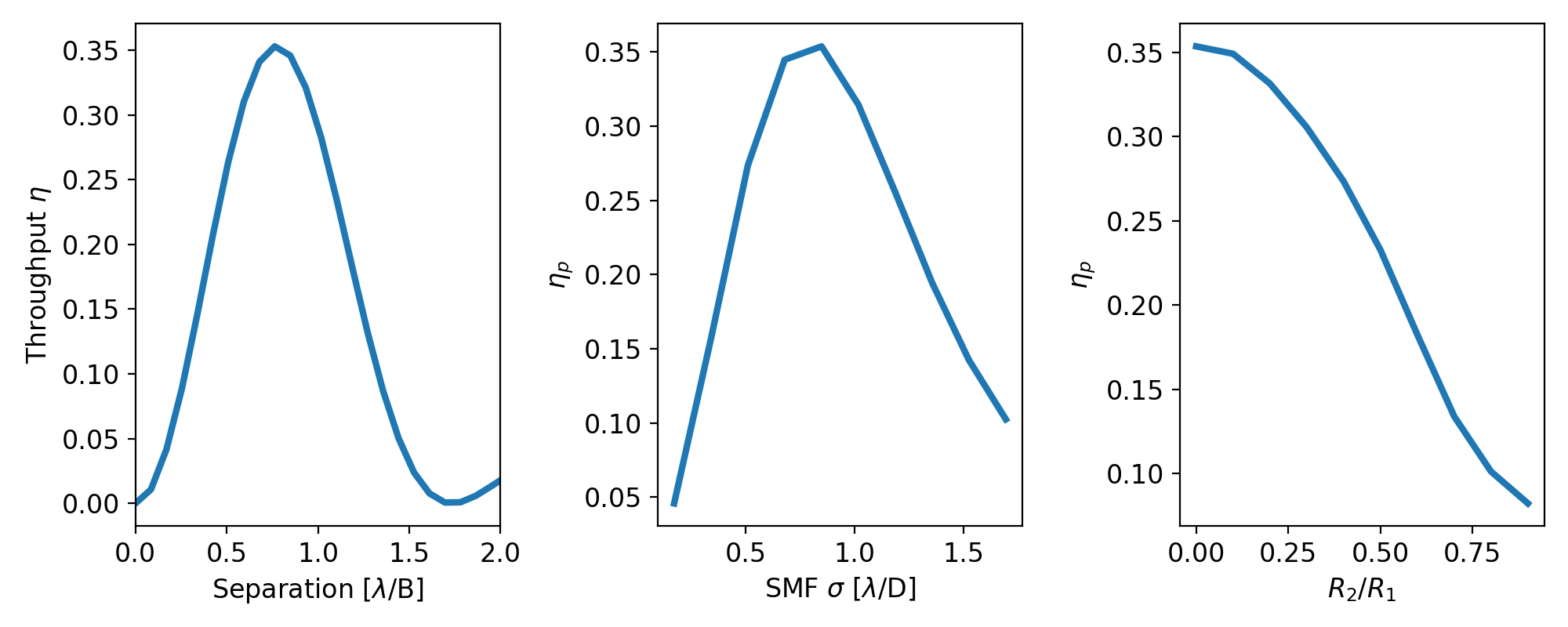}
\caption{{\bf{Left}}: throughput vs. angular separation. {\bf{Middle}}: maximum throughput vs. core size of a single-mode fiber. {\bf{Right}}: peak throughput vs. central obscuration. $R_1$ and $R_2$ are the radius of primary and secondary mirror, respectively.   \label{fig:eta_p_sep}}
\end{figure*}

Without the beam shaping device, the beam shape $\Psi(r, \theta)$ does not fit the Gaussian fundamental mode of a single-mode fiber. Therefore, using a single-mode fiber without the beam shaping device, while not compromising the on-axis starlight suppression, will suffer from a peak coupling efficiency loss. Indeed, our simulation shows that the maximum throughput decreases by 2.3 times if no beam-shaping device is used. 

We also show in Fig. \ref{fig:eta_p_sep} how the peak efficiency is affected by central obscuration of the secondary mirror. For LBT, the radius ratio between the secondary ($R_2$) and the primary mirror ($R_1$) is 0.45 / 4.20 = 0.11. The peak efficiency is reduced by less than 1\%. In addition, the central obscuration does not affect the on-axis starlight suppression as long as the primary and the secondary mirrors are well aligned.

}



\subsection{Gaps in Planet Searching Area}

The area with high coupling efficiency is no longer azimuthally symmetric for the dual-aperture case when compared to the single-aperture case~\citep{Ruane2019}. The implication is that the search time will increase in order to cover all phase angles for a given angular separation. However, for planets with known position angles, follow-up observations can be optimized using the right parallactic angle. 

The key advantage of the dual-aperture case is the spatial resolution that rivals the spatial resolution of ELTs that are coming online in the next decade. We will discuss how we quantify the performance of the dual-aperture FN system in \S \ref{sec:metrics} and lay out science cases that are enabled in \S \ref{sec:sci} by using LBTI as an example. The same concept can also be applied to other dual-aperture interferometers.   

\section{Performance Metrics for a FN system}
\label{sec:metrics}

We use a merit system that is based on the required exposure time to evaluate the performance of a VFN system~\citep{Ruane2018}. {In short, a certain exposure time $\tau$ is required to reach a given signal-to-noise ratio (SNR), where the signal is the photons from the planet, and the noise may come from different sources, e.g., leaked stellar light and thermal background. }

{The total amount of exposure time is a summation of the required exposure times to overcome a variety of noise sources:
\begin{equation}
\tau = \tau_{\rm{L}} + \tau_\Theta + \tau_{\rm{bg}} + \tau_{\rm{dc}} + \tau_{\rm{rd}}, 
\label{eq:t_exp}
\end{equation}
where $\tau_{\rm{L}}$ is the exposure time to reach a given SNR for a specific noise source: the leaked stellar photons due to low-order aberrations. This term is discussed in more details in \S \ref{sec:low_order}. 

The term $\tau_\Theta$ is the required exposure time to overcome the leaked stellar photons due to the finite size of a star. In the case of a circular aperture, this term scales with $D^2$, where $D$ is the aperture size~\citep{Ruane2018}. However, for the interferometer case, $\tau_\Theta$ scales with $D\times B$, where $B$ is the edge-to-edge baseline.

Instrument and sky contribute to thermal background, which requires an exposure time of $\tau_{\rm{bg}}$ to overcome the thermal background noise to reach a given SNR. In this case, background radiance and the solid angle subtended by the fiber determine the thermal background noise that is coupled into the system. 

Detector noise such as the dark current (dc) and readout (rd) noise would require a certain exposure time of $\tau_{\rm{dc}}$ and $\tau_{\rm{rd}}$ to overcome. The relative contribution of the dark current to the leaked stellar photons determines $\tau_{\rm{dc}}$. For $\tau_{\rm{rd}}$, the relevant parameters are detector well depth and readout noise, which determine how many readouts are needed and how much each readout contributes to the noise budget.  } 

All these terms are discussed in detailed in~\citet{Ruane2018}, so we refer readers to \S 3 in that paper. We note in Eq. \ref{eq:t_exp}, $\tau_{\rm{L}}$ absorbs the term $\tau_{\rm{tt}}$ in Eq. 15 in~\citet{Ruane2018} because we consider tip-tilt as low-order aberration. 

{
\subsection{An Example for $\tau$}
Here we use an example to explain how $\tau$ is calculated to overcome a certain noise source. If the dominating noise source is leaked stellar photons, then SNR per spectral channel is derived as follows:
\begin{equation}
{\rm{SNR}} = \frac{S_p}{\sqrt{S_s}} = \frac{\eta_p}{\sqrt{\eta_s}}\frac{\Phi_p}{\sqrt{\Phi_s}}\sqrt{\frac{\tau \lambda A q T}{R}},
\label{eq:t_example}
\end{equation}
where $S$ is photon count and subscripts $p$ and $s$ are for planet and star, respectively. $\eta_p$ is the coupling efficiency at the planet location or the planet throughput, $\eta_s$ is the on-axis throughput or the starlight suppression level. $\Phi_s$ and $\Phi_p$ are star and planet flux in the unit of photons per unit area per unit time per unit wavelength at the primary mirror, $\tau$ is the exposure time, $\lambda$ is the central wavelength, $A$ is the aperture size, $q$ is the quantum efficiency of the detector, $T$ is the instrument throughput that affects the star and the planet equally, and $R$ is the spectral resolution of the spectrograph. 

Solving for $\tau$, we will have:
\begin{equation}
\tau = \frac{\eta_s}{\eta_p^2} \tau_0, 
\label{eq:t_lead_star}
\end{equation}
the required exposure time to achieve a certain SNR in the presence of leaked stellar photons. In the above equation, $\tau_{0}$ is calculated the same way as Eq. 4 in~\citet{Ruane2018}:
\begin{equation}
\tau_{0} = \frac{R}{\lambda}\frac{{\rm{SNR}}^2}{\epsilon^2\Phi_s A q T}, 
\label{eq:t_0}
\end{equation}
where $\rm{SNR}$ is the desired signal to noise ratio and $\epsilon$ is planet-star flux ratio $\Phi_p/\Phi_s$.

}

\subsection{Sensitivity to Low-Order Aberrations}
\label{sec:low_order}
We pay special attention to $\tau_{\rm{L}}$ because it is system-specific. We adopt the same parameterization as~\citet{Ruane2018} except that we change the fixed power-law dependence of 2 to a variable $\gamma$:
\begin{equation}
\tau_{\rm{L}} = \frac{\tau_0}{\eta_p^2} \eta_s \approx \frac{\tau_0}{\eta_p^2} \sum_i (b_i \omega_i)^\gamma.
\label{eq:t_L}
\end{equation}
The term $\eta_s$ can be approximated as the summation of contribution of all Zernike modes. Subscript $i$ is Zernike mode number, $b_i$ is the coefficient describing how starlight suppression depends on Zernike aberration, and $\omega_i$ is the RMS wavefront error (in the unit of 1/$\lambda$) {for the $i_{\rm{th}}$ Zernike mode}.  


We conduct simulation to numerically quantify $b_i$. We use a functional form $\eta_s = (b_i \omega)^\gamma$, where $\gamma=2$ or 4, to fit the numerical points for each Zernike mode. We set one sub-aperture to have zero wavefront error and add aberration to the other sub-aperture. In the case in which both sub-apertures have comparable aberrations, the wavefront error increases by a factor of $\sqrt{2}$.

\begin{figure}[h]
\plotone{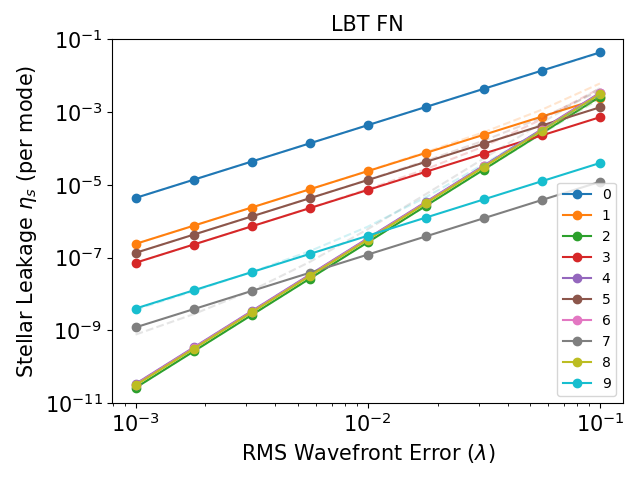}
\caption{Starlight suppression level vs. RMS wavefront error for each Zernike mode for an LBT FN. Solid lines are fitting results, dashed lines connect data points in numerical simulations. Quantitative relationships between $\eta_s$ and wavefront error are given in Table \ref{tab:bi}. \label{fig:eta_s_abb}}
\end{figure}


\subsection{Compared to a Single-Aperture VFN}

The coefficients of sensitivity to low-order aberrations are shown in Fig. \ref{fig:eta_s_abb} and summarized in Table \ref{tab:bi}. Overall, the functional form provides a good approximation. Except for piston, all other even-numbered Zernike modes have a power=4 dependence on aberrations. This is different from the power=2 dependence for VFN~\citep{Ruane2018}. 

The implication is that small aberrations (e.g., RMS wavefront error $<$ $\lambda/100$ per mode) in these Zernike modes contribute negligibly to starlight leakage, and the contribution becomes significant at large aberrations (e.g., RMS wavefront error $\sim$ $\lambda/10$). Because we are interested in the FN performance at small aberrations, these even-numbered Zernike modes can be omitted in $\tau_{\rm{L}}$ calculation (Eq. \ref{eq:t_L}). 


In contrast to the single-aperture VFN~\citep{Ruane2018}, the dual-aperture FN is not azimuthally symmetric, so it loses the advantage of the single-aperture case, i.e., the single-aperture VFN is only sensitive to $Z_n^{\pm1}$ (see Table \ref{tab:bi} for the GMT case and \S \ref{sec:lbt_gmt} for more details). 

In addition, it is also shown in Fig. \ref{fig:eta_s_abb} that the dual-aperture FN is sensitive to piston aberration. This is because changing piston for one sub-aperture while maintaining piston for the other sub-aperture would shift the interferogram along the baseline direction. The effect is similar to the impact of tip-tilt in a single-aperture VFN. 

\begin{deluxetable*}{cccccc}
\tablecaption{Sensitivity of starlight suppression $\eta_s$ to Zernike Aberration, $\eta_s = \sum_i (b_i\omega)^\gamma$, see also Eq. \ref{eq:t_L}. \label{tab:bi}}
\tablewidth{0pt}
\tablehead{
\colhead{OSA Index} & \colhead{Classical Name} & \multicolumn{2}{c}{LBT FN} & \multicolumn{2}{c}{GMT VFN} \\
\colhead{} & \colhead{} & \multicolumn{2}{c}{} & \multicolumn{2}{c}{charge=1} \\
\colhead{} & \colhead{} & \colhead{$b_i$} & \colhead{$\gamma$} & \colhead{$b_i$} & \colhead{$\gamma$}
}
\startdata
00 & Piston & 2.09 & 2 & 0.01 & 2 \\
01 & Tilt & 0.48 & 2 & 2.94 & 2 \\
02 & Tip & 2.26 & 4 & 2.94 & 2 \\
03 & Oblique astigmatism & 0.26 & 2 & 0.01 & 2 \\
04 & Defocus & 2.41 & 4 & 0.01 & 2 \\
05 & Vertical astigmatism & 0.36 & 2 & 0.01 & 2 \\
06 & Vertical trefoil & 2.38 & 4 & 0.01 & 2 \\
07 & Vertical coma & 0.03 & 2 & 2.61 & 2 \\
08 & Horizontal coma & 2.36 & 4 & 2.60 & 2 \\
09 & Oblique trefoil & 0.06 & 2 & 0.01 & 2 
\enddata
\end{deluxetable*}

{

\subsection{The impact of Non-Common Path Aberration (NCPA) on Exposure Time}

AO performance determines the length of exposure time via Eq. \ref{eq:t_L}. For a wavefront sensing system with a non-common path between the sensing channel and the science channel, NCPA becomes the dominant term in the AO error budget. NCPA is $\sim$200 nm for the LBT AO system, dominated by astigmatism followed by $\sim$50 nm trefoil contribution~\citep{Bailey2014}. The 200-nm NCPA corresponds to $\lambda$/10 and $\lambda$/15 for $K$ and $L$ band. 

We believe that the AO performance can be improved to $\lambda$/100 per mode for the following reasons. First, NCPA for the state-of-the-art high-contrast imaging instrument SPHERE is $\sim$50 nm~\citep{NDiaye2013} for stars with $r<9$~\citep{Beuzit2019}. This can be further improved to 10-20 nm with a Zernike phase-mask sensor~\citep{Vigan2019}, which corresponds to better than $\lambda$/100 per mode for $K$ and $L$ band. Furthermore, NCPA can be accounted for by focal-plane wavefront sensing techniques~\citep{Galicher2019}, which shows promising prospects of reducing wavefront error to lower than 10 nm given ample photons~\citep{Bos2019}. We therefore use $\lambda$/100 (or 0.063 rad) per mode as a reasonable baseline RMS wavefront error for subsequent calculations for exposure time. Exposure times for other wavefront error values can be scaled using Eq. \ref{eq:t_L} in the case where optical aberrations dominate the error budget. 

\subsection{Reducing Exposure Time by the Cross-Correlation Technique}

The exposure time from Eq. \ref{eq:t_exp} is based on SNR per spectral channel. In practice, observation covers many spectral channels. Therefore, the final SNR is boosted by combining signals from all spectral channels. This can be done by cross-correlating a template spectrum with the observed spectrum. If the two spectra match, the final SNR may be boosted significantly. The cross-correlation technique has been widely used in characterizing planet atmospheres~\citep[see e.g.,][]{Snellen2010,Snellen2014,Bryan2018}. Since the exposure time scales as the square of SNR (see Eq. \ref{eq:t_0}), the actual exposure time $\tau^{\prime}$ can be reduced by a factor of $\gamma^2$ from the exposure time given by Eq. \ref{eq:t_exp}:
\begin{equation}
\tau^{\prime} = \tau(\tau_0) / \gamma^2, 
\label{eq:t_prime}
\end{equation}
where the boost factor $\gamma$ is:
\begin{equation}
\gamma = \frac{{\rm{SNR_{CCF}}}}{{\rm{SNR_{\delta\lambda}}}}.
\label{eq:gamma}
\end{equation}
${\rm{SNR_{CCF}}}$ is the SNR from the cross-correlation technique and ${\rm{SNR_{\delta\lambda}}}$ is the SNR per spectral channel. 

The boost factor $\gamma$ can be approximated by $\sqrt{\rm{N_{lines}}}$, the number of spectral lines within the spectral coverage~\citep{Snellen2015}. However, this is usually an overly optimistic estimation because of all lines have depths smaller than unity and have finite width. Through numerical simulations,~\citet{Ruane2018} found that $\gamma$ values are 40 and 35 for $K$ and $L$ band. Note that the numerical simulations used a spectrum of an Earth-like planet. However, similar values apply to gas-giant planets because the information content, as quantified by a quality factor~\citep{Bouchy2001}, is similar between an Earth-like planet and a gas giant planet. Planet rotation would reduce the boost factor as the rotation broadens lines and therefore reduces the peak of the cross-correlation function. Numerical simulations suggest that $\gamma$ reduces by $\sim$1.4 from a non-rotating case to a v$\sin i$=15 km$\cdot\rm{s}^{-1}$ case (Otten et al. submitted to A\&A). This corresponds to an increase of exposure time by a factor of 2.  
}

\section{Applications}
\label{sec:sci}

Direct spectroscopy of exoplanets is an alternative way of studying exoplanet atmospheres to transit spectroscopy. Since only less than 10\% of planets transit their host stars, direct spectroscopy in principle makes it more accessible to probe exoplanet atmospheres, especially for the most nearby exoplanets that are detected by the radial velocity technique and do not transit. 


Together with planet mass and metallicity as inferred from radial velocity data, and possibly age from asteroseismology due to their proximity, and planet chemical composition measurements provided by direct spectroscopy, this information can be used as bench marks to test and improve planet atmospheric modeling. This science case is discussed in \S \ref{sec:rv_planet_ground}. 

Dual-aperture FN offers excellent starlight suppression at a spatial resolution that is comparable to that of ELTs (\S \ref{sec:metrics}). Aided by high resolution spectroscopy, the effective starlight suppression level can be improved by another few orders of magnitude~\citep{Wang2017}. This allows us to improve inner working angle (IWA) to observe lower-mass planets that are intrinsically more frequent than gas giant planets that can be currently detected~\citep{Bowler2016,Fernandes2019}. 

Moreover, to alleviate the the large sample size ($\gtrapprox$100) that is usually required to directly image a couple of exoplanets, we can conduct the search for planets around dusty systems, whose long-period planet occurrence rate is boosted by a factor of $\sim$10 compared to systems without such a constraint~\citep{Meshkat2017}. We will discuss this science case in \S \ref{sec:dusty}. 

Direct imaging and spectroscopy of rocky planets in the habitable zone is a major science driver for ELTs and future space missions. Space missions such as HabEx and LUVOIR are limited by spatial resolution $\lambda$/D. Increasing aperture size D will significantly increase the cost. Another cost driver is the cooling systems that are required to reach mid- and thermal-infrared wavelengths, which the above space mission avoids. However, avoiding long wavelengths in infrared will limit these space missions' ability in searching for biosignatures, which usually have more much abundant spectral lines in infrared than at shorter wavelengths. Space interferometry creates a niche in high-spatial-resolution infrared spectroscopy for temperate planets around nearby M stars, which are traditionally in the reign of ground-based ELTs. This science case will be discussed in \S \ref{sec:space_inter}.   

\subsection{Follow-up Observations of Exoplanets Detected by other techniques}
\label{sec:rv_planet_ground}

\begin{figure}[h]
\plotone{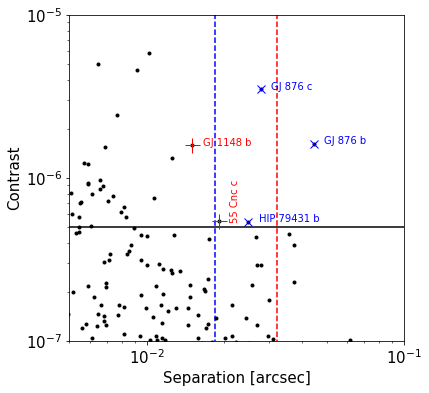}
\caption{Planet-star contrast vs. angular separation. Colored data points have angular separation larger than 15 mas and planet-star contrast higher than $5\times10^{-7}$ and therefore amenable for direct spectroscopy (see also Table \ref{tab:rv_followup}). Red pluses are targets in the north and blue crosses are targets in the south. Blue and red dashed lines mark 1 $\lambda$/B for $K$ and $L$ band for a baseline of 22.8 meter.     \label{fig:rv_followup}}
\end{figure}

We use 4152 exoplanets from the NASA Exoplanet Archive (NEA) service\footnote{\url{https://exoplanetarchive.ipac.caltech.edu/}}. We put these planets on a separation - planet-star contrast plot as shown in Fig. \ref{fig:rv_followup} in order to select amenable targets. Targets with contrast lower than $5\times10^{-7}$ and angular separation larger than 15 mas are given in Table \ref{tab:rv_followup}. We focus on $K$ and $L$ band, which are a trade-off between thermal background and wavefront aberration. As an example, we use the LBTI to present the following two science cases. Below we detail how separation and planet-star contrast are calculated based on information available from NEA. 

We calculate the planet-star separation based on their reported distance and semi-major axis. When the latter is not available, we calculate it using orbital period and stellar mass. Planet-star contrast $\epsilon$ is calculated using the following equation: 
\begin{equation}
\epsilon = \left(\frac{R_p}{a}\right)^2 \times A_g, 
\label{eq:contrast}
\end{equation}
where, $R_p$ is planet radius, $a$ is semi-major axis, and $A_g$ is albedo, which is assumed to be 0.3. {This roughly corresponds to the albedo of Jupiter at a 60-degree phase angle with a Lambertian phase function.} When planet radius is not available, we use the mass-radius relation in~\citet{Thorngren2019} to calculate radius for planet masses between 15 M$_\oplus$ and 12 M$_{\rm{Jupiter}}$. For planet masses outside the range, we use the mass-radius relation in~\citet{Chen2017}. {For RV-detected planets, the orbital inclination with respect to the sky-plane is unknown. In this case, we assume an edge-on orbit with $i=90$ degree.}

The code for target selection and exposure time calculation is available through a Python notebook on \faicon{github}GitHub\footnote{\url{https://github.com/wj198414/VFN}}. Here we present two examples to illustrate how to interpret the calculated exposure times given in Table \ref{tab:rv_followup}. 

\subsubsection{55 Cnc c in $K$-band LBTI observation}
55 Cnc c~\citep{McArthur2004} is among the most challenging exoplanets on our list in terms angular separation (19.2 mas) and planet-star contrast ($7.18\times10^{-7}$). The angular separation corresponds to 1.04 $\lambda/B$ in $K$ band assuming a 22.8-m baseline. 

The details of our simulation are provided in Table \ref{tab:55cnc_c_k}. Planet throughput is calculated based on the dependence on angular separation (Fig. \ref{fig:eta_p_sep}). {We assume an RMS error of $\lambda/100$ per mode, which translates to a starlight suppression level of $4.81\times10^{-4}$ based on the expression of $\eta_s$ in  Eq. \ref{eq:t_L}. We have the following numbers for required exposure times to overcome various noise sources (\S \ref{sec:metrics}): $\tau_{\rm{L}}=2.70\times10^{7}$ s, $\tau_{\Phi}=6.24\times10^{5}$ s, $\tau_{\rm{bg}}=2.94\times10^{4}$ s, $\tau_{\rm{dc}}=1.65\times10^{5}$ s, and $\tau_{\rm{rd}}=1.65\times10^{3}$ s. The summation of the above terms is $2.78\times10^{7}$ s, or 7739 hours. Taking into account the boost factor $\gamma$ thanks to the cross-correlation technique, the required exposure time is reduced to $7739 / 40^2 = 4.84$ hours for a 5-$\sigma$ detection. }

\subsubsection{55 Cnc c in $L$-band LBTI observation}

{The angular separation corresponds to 0.59 $\lambda/B$ in $L$ band.} Planet throughput is assumed to be 31.0\% for a 0.59 $\lambda$/B angular separation (see Fig. \ref{fig:eta_p_sep}). Star suppression level is again  $4.81\times10^{-4}$ if assuming $\lambda/100$ per mode wavefront error. However, $L$-band wavefront quality is better than that of $K$ band, so $\eta_s$ should be lower. A full description of the parameters used in the simulation are given in Table. \ref{tab:55cnc_c_l}.   

{We have the following numbers for required exposure times to overcome various noise sources (\S \ref{sec:metrics}): $\tau_{\rm{L}}=3.16\times10^{7}$ s, $\tau_{\Phi}=2.38\times10^{5}$ s, $\tau_{\rm{bg}}=1.25\times10^{9}$ s (assuming a $L$ band thermal background of 2.0 mag per square arcsec), $\tau_{\rm{dc}}=2.98\times10^{5}$ s, and $\tau_{\rm{rd}}=1.93\times10^{3}$ s. The summation of the above terms is $128\times10^{9}$ s, or 356780 hours. The required exposure time is therefore limited by the $L$-band thermal background noise. Accounting for the boost factor of $\sim$35 that is brought by the cross-correlation technique, the required exposure time is reduced to 291 hours. }

\begin{deluxetable*}{cccccccccccc}
\tablecaption{A list of gas giant planets that are amenable for direct spectroscopy (see also Fig. \ref{fig:rv_followup}). \label{tab:rv_followup}}
\tabletypesize{\scriptsize}
\tablewidth{0pt}
\tablehead{
\colhead{} & \colhead{} & \colhead{} & \colhead{} & \colhead{} & \colhead{} & \colhead{} & \colhead{} & \multicolumn{2}{c}{{$K$}} & \multicolumn{2}{c}{{$L$}} \\
\colhead{Planet name} & \colhead{Distance} & \colhead{$K$} & \colhead{$L$} & \colhead{a} & \colhead{$R_p$} & \colhead{Contrast} & \colhead{Sep.} & \colhead{Sep.} & \colhead{t$^{\prime,\dagger}$} & \colhead{Sep.} & \colhead{t$^{\prime,\dagger}$} \\
\colhead{} & \colhead{pc} & \colhead{mag} & \colhead{mag} & \colhead{A.U.} & \colhead{$R_{\rm{Jupiter}}$} & \colhead{} & \colhead{mas} & \colhead{$\lambda/D$} & \colhead{hour} & \colhead{$\lambda/D$} & \colhead{hour}
}
\startdata
55 Cnc c             &   12.6 &   4.0 & 4.0 &  0.241 &   0.68 & 5.47e-07 &   19.2 &  1.04 &     4.84 & 0.59 & 291.25 \\
GJ 1148 b            &   11.0 &   6.8 & 6.7 &  0.166 &   0.80 & 1.58e-06 &   15.1 &  0.82 &     4.89 & 0.47 & 8042.00\\
GJ 876 b             &    4.7 &   5.0 & 4.9 &  0.208 &   1.01 & 1.61e-06 &   44.5 &  2.42 &   105.82 & 1.38 & 2667.70\\
GJ 876 c             &    4.7 &   5.0 & 4.9 &  0.130 &   0.93 & 3.49e-06 &   27.7 &  1.50 &    22.27 & 0.86 & 28.30\\
HIP 79431 b          &   14.5 &   6.6 & 6.5 &  0.360 &   1.01 & 5.36e-07 &   24.8 &  1.34 & 468.98 & 0.77 & 23277.22\\
\enddata
\tablecomments{$\dagger$: Dual-aperture FN needs two pointing positions for covering areas around 1 $\lambda$/B. Doubling exposure time is not accounted here.}
\end{deluxetable*}

\subsection{Direct Spectroscopy of Exoplanets Embedded in Systems with Disks}
\label{sec:dusty}

We select targets using the Catalog of Circulstellar Disks\footnote{\url{https://webdisks.jpl.nasa.gov/}}. There are 48 debris disk systems with $r$ magnitudes brighter than 8th and inclinations lower than 60 degree, i.e., more face-on systems (Table \ref{tab:disk_sys}). The magnitude cut is to ensure optimal AO performance and the inclination cut is to minimize the extinction due to the increasing viewing angle. Below we use HD 104860, the faintest debris-disk system in our sample with $R=8.0$, as an example to demonstrate the feasibility of using LBTI and the dual-aperture FN to search for planets in debris-disk systems. We provide a Python notebook to compute required exposure time to achieve the sensitivity for a given planet-star contrast.

\subsubsection{HD 104860}

HD 104860~\citep[][and references therein]{Morales2013} represents the worst-case scenario among all targets because it is the faintest debris-disk system in our sample with $R = 8.0$. We convert $R$-band magnitude into $K$ or $L$-band magnitude using the updated Table 5 in~\citet{Pecaut2013} for a given effective temperature. In calculating exposure times, we set planet-star contrast to $10^{-6}$. While the contrast is comparable to the state-of-the-art performance, the greatest gain is the IWA of the FN, which brings the IWA to $\sim$20 mas (Table \ref{tab:hd104960_k}). 

The breakdown of the required exposure times to overcome various noise sources (\S \ref{sec:metrics}) are as follows: $\tau_{\rm{L}}=2.31\times10^{8}$ s, {$\tau_{\Phi}=4.18\times10^{5}$ s}, $\tau_{\rm{bg}}=3.97\times10^{6}$ s, $\tau_{\rm{dc}}=2.22\times10^{7}$ s, and $\tau_{\rm{rd}}=1.41\times10^{4}$ s. The total exposure time is $2.59\times10^{8}$ s, or 71963 hours for $K$-band observation. Accounting for the boost factor of $\sim$40 that is brought by the cross-correlation technique, the required exposure time is reduced to 44.9 hours. 

We note that the final exposure time is very sensitive to planet-star contrast. Relaxing the targeted planet-star contrast by two times would reduce the exposure time by a factor of 4 (Eq. \ref{eq:t_0}). The exposure time is also sensitive to planet throughput and wavefront error to the second power (Eq. \ref{eq:t_L}). Therefore, improving wavefront quality and planet throughput is the key to increase the efficiency of planet search. 

\subsubsection{Background induced by disk brightness}

Below we will show that the background noise due to the emissivity of sky and instrument is almost always higher than the background due to the disk brightness. We can therefore only consider the sky and instrument background when calculating $\tau_{\rm{bg}}$. Using HD 191089~\citep{Soummer2014} as an example, the system has a bright debris disk that has a flux of 1 mJy per square arcsec~\citep{Ren2019} in $H$ band. In comparison, the star is 3750 mJy in $H$ band. Given the extent of the disk at $\sim$1 square arcsec, the ratio between the integrated disk flux and the star flux is $2.6\times10^{-4}$ or a delta magnitude of 8.9 mag. Since scattered light is the major component in near infrared, it is reasonable to assume that the ratio is similar in $K$ and $L$ band. In the case of HD 191089, the background induced by disk brightness is $\sim$14 mag per square arcsec, lower than the assumed thermal background in our calculation, i.e., 12.2 mag per square arcsec in $K$ band and  2.0 mag per square arcsec in $L$ band.  

\begin{deluxetable*}{cccccc}
\tablecaption{A list of nearby debris disk systems with inclinations lower than 60 degrees. \label{tab:disk_sys}}
\tabletypesize{\scriptsize}
\tablewidth{0pt}
\tablehead{
\colhead{Count} & \colhead{Star} & \colhead{RA} & \colhead{Dec} & \colhead{$R$} & \colhead{Distance} \\
\colhead{} & \colhead{} & \colhead{hh mm ss} & \colhead{dd mm ss} & \colhead{mag} & \colhead{pc} 
}
\startdata
 1 & 99 Her               & 18 07 01.54          & +30 33 43.7          &    4.7 &   15.6 \\
 2 & AB Aur               & 04 55 45.93          & +30 33 03.6          &    7.1 &  144.0 \\
 3 & beta Leo             & 11 49 03.58          & +14 34 19.4          &    2.0 &   11.1 \\
 4 & beta Tri             & 02 09 32.63          & +34 59 14.3          &    2.9 &   38.9 \\
 5 & epsilon Eri          & 03 32 55.84          & -09 27 29.7          &    3.8 &    3.2 \\
 6 & eta Crv              & 12 32 04.23          & -16 11 45.6          &    4.4 &   18.2 \\
 7 & gamma Oph            & 17 47 53.56          & +02 42 26.2          &    3.7 &   29.1 \\
 8 & HD 100453            & 11 33 05.57          & -54 19 28.5          &    7.8 &  103.0 \\
 9 & HD 100546            & 11 33 25.44          & -70 11 41.2          &    6.7 &  103.0 \\
10 & HD 104860            & 12 04 33.73          & +66 20 11.7          &    8.0 &   47.9 \\
11 & HD 10647             & 01 42 29.32          & -53 44 27.0          &    5.5 &   17.4 \\
12 & HD 107146            & 12 19 06.50          & +16 32 53.9          &    6.7 &   27.5 \\
13 & HD 10939             & 01 46 06.26          & -53 31 19.3          &    5.1 &   57.0 \\
14 & HD 110058            & 12 39 46.20          & -49 11 55.5          &    8.0 &  107.0 \\
15 & HD 138813            & 15 35 16.11          & -25 44 03.0          &    7.3 &  151.0 \\
16 & HD 138965            & 15 40 11.56          & -70 13 40.4          &    6.5 &   77.3 \\
17 & HD 141378            & 15 48 56.80          & -03 49 06.6          &    5.6 &   49.2 \\
18 & HD 141569A           & 15 49 57.76          & -03 55 16.2          &    7.1 &   99.0 \\
19 & HD 153053            & 17 00 06.28          & -54 35 49.8          &    5.7 &   50.7 \\
20 & HD 156623            & 17 20 50.62          & -45 25 14.5          &    7.3 &  118.0 \\
21 & HD 15745             & 02 32 55.81          & +37 20 01.4          &    7.5 &   64.0 \\
22 & HD 159492            & 17 38 05.52          & -54 30 01.6          &    5.3 &   42.2 \\
23 & HD 163296            & 17 56 21.26          & -21 57 21.6          &    6.9 &  101.0 \\
24 & HD 166               & 00 06 36.78          & +29 01 17.4          &    5.6 &   13.7 \\
25 & HD 16743             & 02 39 07.56          & -52 56 05.3          &    6.9 &   58.9 \\
26 & HD 170773            & 18 33 00.92          & -39 53 31.3          &    6.7 &   37.0 \\
27 & HD 172555            & 18 45 26.90          & -64 52 16.5          &    4.9 &   29.2 \\
28 & HD 181327            & 19 22 58.94          & -54 32 17.0          &    7.1 &   50.6 \\
29 & HD 183324            & 19 29 00.99          & +01 57 01.6          &    5.9 &   59.0 \\
30 & HD 188228            & 20 00 35.56          & -72 54 37.8          &    4.0 &   32.2 \\
31 & HD 191089            & 20 09 05.21          & -26 13 26.5          &    7.0 &   54.0 \\
32 & HD 195627            & 20 35 34.85          & -60 34 54.3          &    4.8 &   27.8 \\
33 & HD 206893            & 21 45 21.90          & -12 47 00.1          &    7.1 &   38.3 \\
34 & HD 20794             & 03 19 55.65          & -43 04 11.2          &    3.7 &    6.0 \\
35 & HD 21997             & 03 31 53.65          & -25 36 50.9          &    6.5 &   71.9 \\
36 & HD 30422             & 04 46 25.75          & -28 05 14.8          &    6.3 &   57.5 \\
37 & HD 38858             & 05 48 34.94          & -04 05 40.7          &    5.4 &   15.2 \\
38 & HD 53143             & 06 59 59.66          & -61 20 10.3          &    6.9 &   18.4 \\
39 & HD 71155             & 08 25 39.63          & -03 54 23.1          &    3.9 &   37.5 \\
40 & HD 74873             & 08 46 56.02          & +12 06 35.8          &    7.2 &   61.0 \\
41 & HD 95086             & 10 57 03.02          & -68 40 02.5          &    7.5 &   90.4 \\
42 & HR 8799              & 23 07 28.71          & +21 08 03.3          &    5.8 &   40.0 \\
43 & Kappa CrB            & 15 51 13.93          & +35 39 26.6          &    4.1 &   31.1 \\
44 & lambda Boo           & 14 16 23.02          & +46 05 17.9          &    4.1 &   30.3 \\
45 & MWC 480              & 04 58 46.26          & +29 50 37.0          &    7.8 &  161.0 \\
46 & Tau Ceti             & 01 44 04.08          & -15 56 14.9          &    3.5 &    3.6 \\
47 & Vega                 & 18 36 56.34          & +38 47 01.3          &    0.1 &    7.8 \\
48 & zeta Lep             & 05 46 57.34          & -14 49 19.0          &    3.4 &   21.6 \\
\enddata
\end{deluxetable*}

\subsection{Characterizing Rocky Planets Around M Stars with Space Interferometric Array}
\label{sec:space_inter}

The small angular separations ($<$20 mas) of habitable planets around M stars are formidable for space direct-imaging missions due to limited aperture sizes. Moreover, searching for multiple tracers of biosignatures (e.g., water, oxygen, and methane), which reduces the likelihood of false positives~\citep{Goldman2014, Harman2015}, requires infrared observations. Observing at infrared wavelengths further decreases the spatial resolution for space missions. 

Infrared interferometry provides a solution to the above issue~\citep[e.g., ][]{Kammerer2018}. In addition, the dual-aperture FN concept alleviates many of the technical challenges towards a space interferometry mission~\citep{Monnier2019}. 

\begin{figure}[h]
\plotone{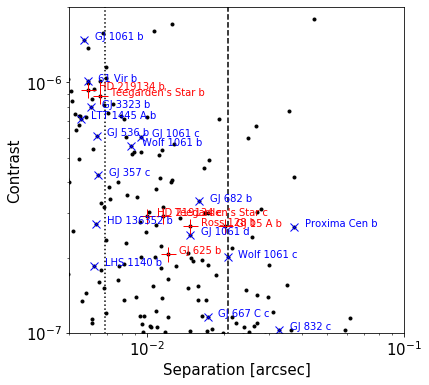}
\caption{Planet-star contrast vs. angular separation. Colored data points have angular separation larger than 5 mas and planet-star contrast higher than $1\times10^{-7}$ and therefore amenable for direct spectroscopy with a space interferometric mission (see also Table \ref{tab:space_rv}). Red pluses are targets in the north and blue crosses are targets in the south. Dotted and dashed lines mark 1 $\lambda$/B at 1 $\mu$m for a baseline of 30 and 10 meter.  \label{fig:space_dual_ap}}
\end{figure}

We again start from 4152 exoplanets from the NASA Exoplanet Archive service. Following the angular separation and planet-star contrast calculations that are detailed in \S \ref{sec:rv_planet_ground}, we select planets with (1) contrasts lower than $1\times10^{-7}$; (2) angular separations larger than 5 mas; and (3) radii smaller than 0.2 $R_{\rm{Jupiter}}$. Table \ref{tab:space_rv} lists and Fig. \ref{fig:space_dual_ap} shows the 23 potential rocky planets that meet the above criteria. We provide two examples below for the purpose of feasibility demonstration. Calculations for other planets are available through a Python notebook that is available on GitHub\footnote{\url{https://github.com/wj198414/VFN}}. 





\subsubsection{GJ 1061 b}

While GJ 1061 b~\citep{Dreizler2020} has a favorable planet-star contrast at $1.47\times10^{-6}$, its angular separation (5.7 mas) poses challenges for direct spectroscopy. In the following calculation, we assume a sub-aperture diameter of 4 meter and a baseline of 50 meter. This corresponds to an angular separation of 0.69 $\lambda$/B for $K$ band at the 50-meter baseline. Wavefront RMS error is $\lambda$/100 per mode and this translates to a star suppression level of $4.81\times10^{-4}$. For thermal background, we assume a level that is comparable to JWST thermal background at 0.2 MJy/SR, which is 20.4 mag per square arcsec\footnote{\url{http://ssc.spitzer.caltech.edu/warmmission/propkit/pet/magtojy/}}. Full parameters in simulation are given in Table \ref{tab:gj1061b_k}. 

The breakdown of the required exposure times to overcome various noise sources (\S \ref{sec:metrics}) are as follows: $\tau_{\rm{L}}=1.06\times10^{8}$ s, {$\tau_{\Phi}=8.87\times10^{5}$ s}, $\tau_{\rm{bg}}=6.22\times10^{2}$ s, $\tau_{\rm{dc}}=3.10\times10^{7}$ s, and $\tau_{\rm{rd}}=6.47\times10^{3}$ s. Adding up these terms leads to a total exposure time of $1.38\times10^{8}$ s, or 38363 hours for a $K$-band observation. Accounting for the boost factor of $\sim$40 that is brought by the cross-correlation technique, the required exposure time is reduced to 23.9 hours. 

{The total exposure time has two major contributions with the same order of magnitude: the low-order aberration and the dark current. The low-order aberration component can be improved by reducing wavefront error. For reference, $\sim$20 nm RMS error (i.e., $<\lambda$/100 per mode for $K$-band and redder wavelengths) is at the level of JWST wavefront error~\citep{Aronstein2016}. The dark current component becomes significant because of the increasing ratio between dark current to stellar flux due to the decreasing aperture size~\citep[see Eq. 13 in][]{Ruane2018}.
}


\subsubsection{Proxima Cen b}

Proxima Cen b~\citep{Anglada2016} is the closest planetary system to the solar system and therefore presents a compelling case for direct spectroscopy. Here we discuss a case for an $L$-band observation with space-based dual-aperture FN (Table \ref{tab:pro_cen_b_l}).

There are two game changers for the space-based observation. First, the thermal background, which is the major limitation for ground-based observations, is significantly reduced. We assume a JWST thermal background level at 0.2 MJy/SR, or 19.5 mag per square acrsec. Second, space interferometry achieves a superior spatial resolution to any previous space missions that allows Proxima Cen b to be observed at 1.03 $\lambda$/B at a 20-meter baseline in $L$ band.

The breakdown of the required exposure times to overcome various noise sources (\S \ref{sec:metrics}) are as follows: $\tau_{\rm{L}}=6.64\times10^{8}$ s, {$\tau_{\Phi}=4.42\times10^{6}$ s}, $\tau_{\rm{bg}}=8.78\times10^{3}$ s, $\tau_{\rm{dc}}=3.93\times10^{7}$ s, and $\tau_{\rm{rd}}=4.05\times10^{4}$ s. Adding up these terms leads to a total exposure time of $7.07\times10^{8}$ s, or 196660 hours for $L$-band observation. Accounting for the boost factor of $\sim$35 that is brought by the cross-correlation technique, the required exposure time is reduced to 160.5 hours. 

The major limiting factor for the total required exposure time is the low-order aberration. Note that the exposure time is comparable to numbers from HDC simulations for ELT ground-based observations~\citep{Wang2017}. 

\begin{deluxetable*}{cccccccccc}
\tablecaption{A list of nearby planets around M stars that are amenable for direct spectroscopy (see also Fig. \ref{fig:space_dual_ap}). \label{tab:space_rv}}
\tablewidth{0pt}
\tablehead{
\colhead{Count} & \colhead{Star} & \colhead{RA} & \colhead{Dec} & \colhead{Distance} & \colhead{$K$} & \colhead{a} & \colhead{$R_p$} & \colhead{Contrast} & \colhead{Separation} \\
\colhead{} & \colhead{} & \colhead{deg} & \colhead{deg} & \colhead{pc} & \colhead{mag} & \colhead{A.U.} & \colhead{$R_{\rm{Jupiter}}$} & \colhead{} & \colhead{mas} 
}
\startdata
 1 & 61 Vir b             &   199.601318 &   -18.311195 &    8.5 &   3.0 &  0.050 &   0.19 & 1.01e-06 &    5.9 \\
 2 & GJ 1061 b            &    53.998836 &   -44.512634 &    3.7 &   6.6 &  0.021 &   0.10 & 1.47e-06 &    5.7 \\
 3 & GJ 1061 c            &    53.998836 &   -44.512634 &    3.7 &   6.6 &  0.035 &   0.10 & 6.07e-07 &    9.5 \\
 4 & GJ 1061 d            &    53.998836 &   -44.512634 &    3.7 &   6.6 &  0.054 &   0.10 & 2.47e-07 &   14.7 \\
 5 & GJ 15 A b            &     4.595356 &    44.022953 &    3.6 &   4.0 &  0.072 &   0.14 & 2.67e-07 &   20.2 \\
 6 & GJ 3323 b            &    75.489280 &    -6.946263 &    5.4 &   6.7 &  0.033 &   0.11 & 7.96e-07 &    6.1 \\
 7 & GJ 357 c             &   144.006821 &   -21.660797 &    9.4 &   6.5 &  0.061 &   0.15 & 4.26e-07 &    6.5 \\
 8 & GJ 536 b             &   210.263290 &    -2.654864 &   10.4 &   5.7 &  0.067 &   0.20 & 6.11e-07 &    6.4 \\
 9 & GJ 625 b             &   246.352600 &    54.304104 &    6.5 &   5.8 &  0.078 &   0.14 & 2.07e-07 &   12.1 \\
10 & GJ 667 C c           &   259.745085 &   -34.996827 &    7.2 &   6.0 &  0.125 &   0.16 & 1.16e-07 &   17.2 \\
11 & GJ 682 b             &   264.265259 &   -44.319214 &    5.0 &   5.6 &  0.080 &   0.18 & 3.35e-07 &   16.0 \\
12 & GJ 832 c             &   323.391571 &   -49.009006 &    5.0 &   4.5 &  0.163 &   0.20 & 1.03e-07 &   32.8 \\
13 & HD 136352 b          &   230.450623 &   -48.317627 &   14.7 &   4.2 &  0.093 &   0.19 & 2.73e-07 &    6.4 \\
14 & HD 219134 b          &   348.320740 &    57.168354 &    6.5 &   3.3 &  0.039 &   0.14 & 9.33e-07 &    5.9 \\
15 & HD 219134 c          &   348.320740 &    57.168354 &    6.5 &   3.3 &  0.065 &   0.14 & 2.93e-07 &   10.0 \\
16 & LHS 1140 b           &    11.247240 &   -15.271532 &   15.0 &   8.8 &  0.094 &   0.15 & 1.85e-07 &    6.2 \\
17 & LTT 1445 A b         &    45.464111 &   -16.593372 &    6.9 &   6.5 &  0.038 &   0.12 & 7.15e-07 &    5.5 \\
18 & Proxima Cen b        &   217.428955 &   -62.679485 &    1.3 &   4.4 &  0.049 &   0.10 & 2.65e-07 &   37.3 \\
19 & Ross 128 b           &   176.934982 &     0.804563 &    3.4 &   5.7 &  0.050 &   0.10 & 2.67e-07 &   14.7 \\
20 & Teegarden's Star b   &    43.253708 &    16.881289 &    3.8 &   7.6 &  0.025 &   0.09 & 8.82e-07 &    6.6 \\
21 & Teegarden's Star c   &    43.253708 &    16.881289 &    3.8 &   7.6 &  0.044 &   0.09 & 2.94e-07 &   11.6 \\
22 & Wolf 1061 b          &   247.575241 &   -12.662594 &    4.3 &   5.1 &  0.037 &   0.11 & 5.57e-07 &    8.7 \\
23 & Wolf 1061 c          &   247.575241 &   -12.662594 &    4.3 &   5.1 &  0.089 &   0.15 & 2.01e-07 &   20.6 \\
\enddata
\end{deluxetable*}

\section{Comparing FN and VFN}
\label{sec:fn_vfn}
\subsection{The Connection}
Both a FN and a VFN are a nuller, i.e., a device that suppresses on-axis starlight by manipulating the phase of an EM field. A FN achieves the phase manipulation by changing piston. A VFN achieves the phase manipulation with a vortex plate. For an azimuthally-changing EM, a FN and a VFN deliver a similar performance in terms of starlight suppression and IWA.  

\begin{figure}[h]
\plotone{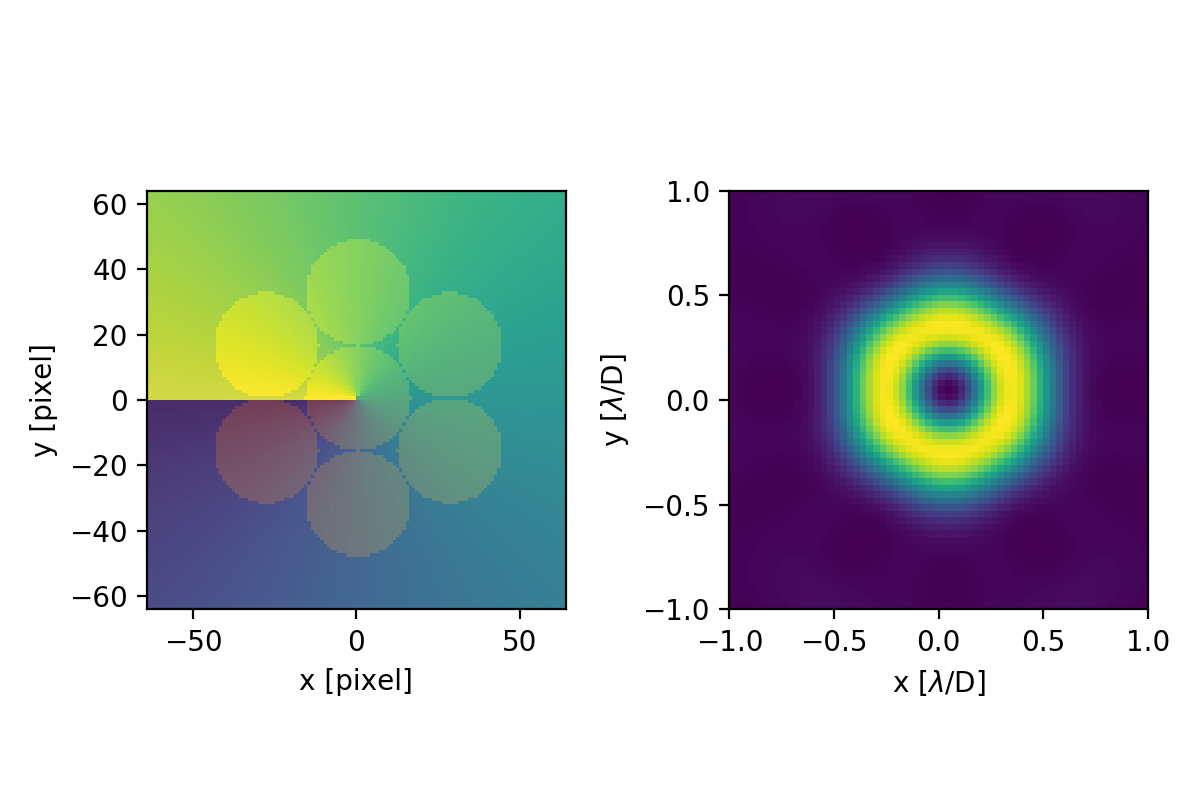}
\caption{{\bf{Left}}: 7 sub-apertures of GMT. The phase ramp of a charge=1 vortex plate is also overplotted. {\bf{Right}}: coupling map of a charge=1 VFN at GMT.     \label{fig:gmt_vfn_1}}
\end{figure}

\begin{figure}[h]
\plotone{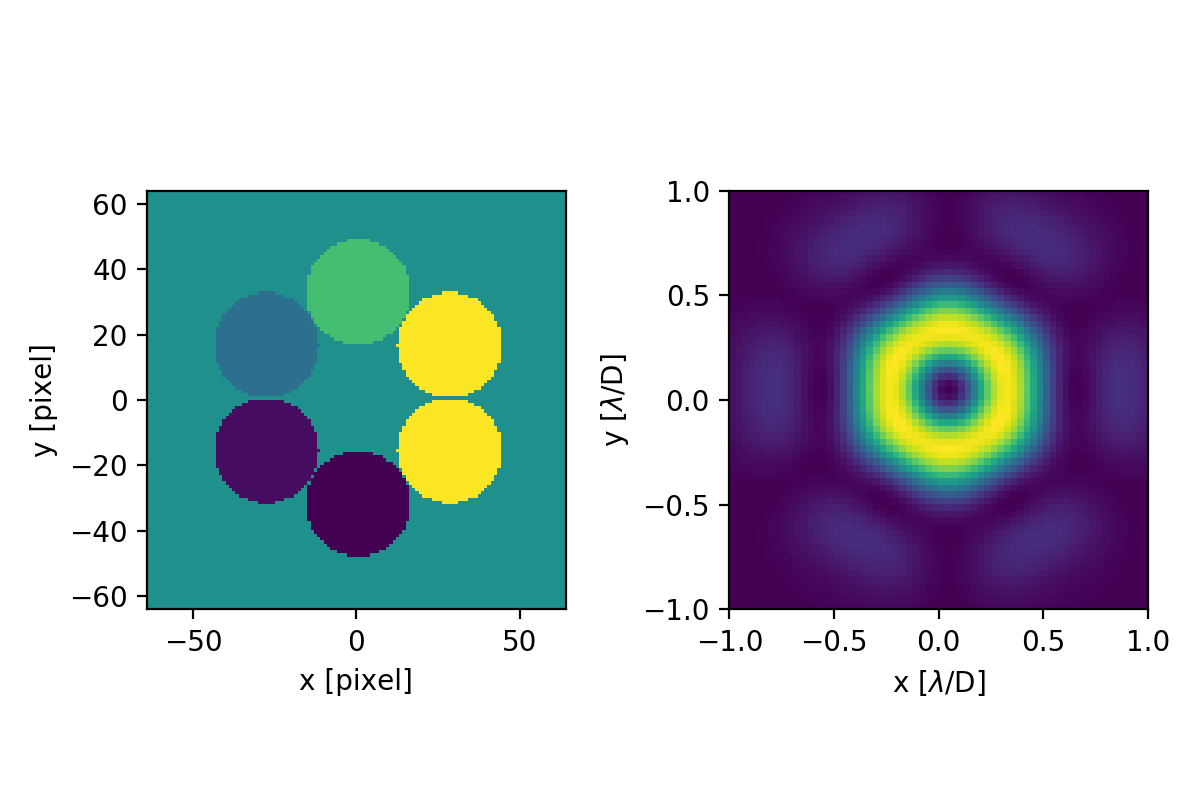}
\caption{{\bf{Left}}: The six outer sub-apertures of GMT with phases changing from 0 to 5$\pi$/6 with an increment of $\pi$/6. {\bf{Right}}: coupling map of a FN at GMT.     \label{fig:gmt_fn_1}}
\end{figure}

We use the Giant Magellan Telescope~\citep[GMT, ][]{Johns2012} configuration to illustrate the similarity between a FN and a VFN. For the VFN setup, we use a vortex plate with charge=1. The resulting coupling map is shown in Fig. \ref{fig:gmt_vfn_1}. For the FN setup, we block the central sub-aperture and use only the outer six sub-apertures. To achieve a similar performance to the GMT VFN, we change piston for the six sub-apertures so that their phases change from 0 to 5$\pi$/6 with an increment of $\pi$/6. This effectively create a charge=1 phase ramp. The resulting coupling map of the GMT FN setup is similar to that of the GMT VFN setup (Fig. \ref{fig:gmt_fn_1}). 

Note that the very same idea can be applied to other ELTs. In addition to changing piston and using a vortex plate, phase manipulation can also be achieved with a deformable mirror. 

\subsection{IWA for a FN with a phase ramp}

\begin{figure}[h]
\plotone{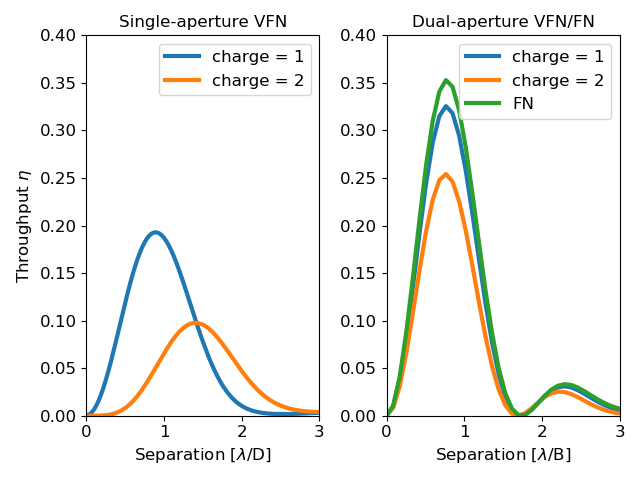}
\caption{{\bf{Left}}: planet throughput vs. angular separation for a single-aperture VFN. {\bf{Right}}: planet throughput vs. angular separation for a dual-aperture VFN or FN.     \label{fig:single_dual_aperture_vfn_iwa}}
\end{figure}

\begin{figure}[h]
\plotone{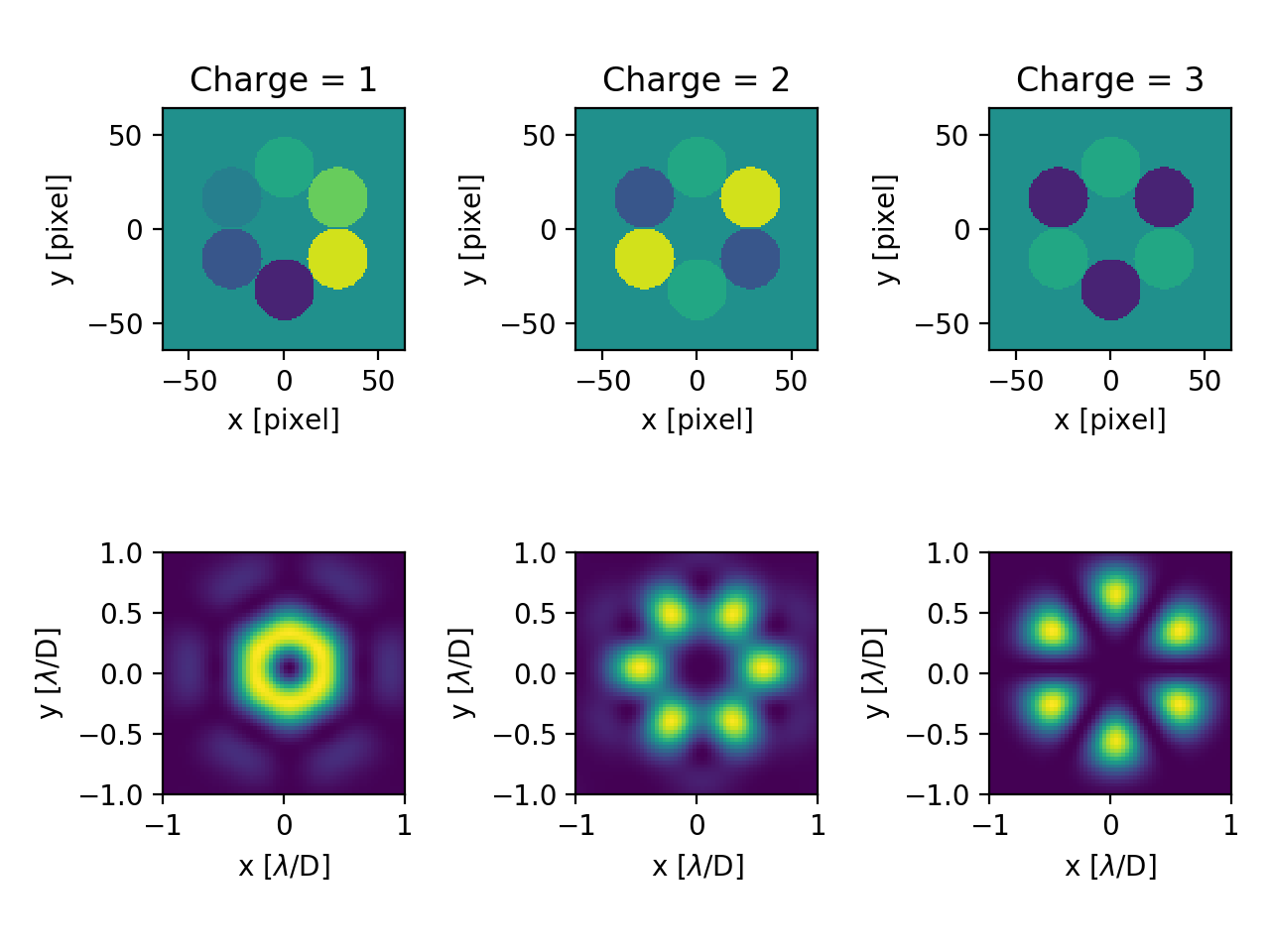}
\caption{{\bf{Left Column}}: top panel is the phase ramp of a FN for GMT. Phases of the six outer sub-apertures change from 0 to 5$\pi$/6 with an increment of $\pi$/6. This correspond to charge = 1. Bottom panel is the corresponding coupling map. {\bf{Middle Column}}: the same as the left column except charge = 2, i.e., phases of the six outer sub-apertures change from 0 to 5$\pi$/3 with an increment of $\pi$/3. {\bf{Right Column}}: the same as the left column except charge = 3.   \label{fig:gmt_fn_123}}
\end{figure}

IWA for a VFN (in the unit of $\lambda$/D) is determined by the charge number of a vortex plate~\citep{Ruane2018}, see also Fig. \ref{fig:single_dual_aperture_vfn_iwa}. For a FN, the IWA as a function of charge number follows the same trend. Three examples are given in Fig. \ref{fig:gmt_fn_123} to illustrate the effect of charge number on IWA. 

As the effective charge number increases, the IWA of a FN is pushed outward. This is quantitatively similar to a VFN on GMT (Fig. \ref{fig:single_dual_aperture_vfn_iwa} Left). However, one noticeable difference is the searching area (bottom rows of Fig. \ref{fig:gmt_fn_123}). Unlike a VFN, the FNs with charge number of 2 and 3 have a partial coverage at a given $\lambda$/D.  

Since the IWA of a VFN or a FN with an effective phase ramp is determined by the charge number, there is a trade-ff between the system complexity and a full coverage at a given angular separation. For a VFN, the system is more complicated with the addition of a vortex plate, but the coverage is continuous for a given $\lambda$/D. In contrast, a FN system is simpler since phase is controlled by the piston of each sub-aperture, but there are insensitive planet-search areas along an annulus. However, this drawback of a FN can be compensated by the simplicity/flexibility of the system: by combining different configurations, e.g., adding up coupling maps for charge 1, 2, and 3 as shown in Fig. \ref{fig:gmt_fn_123}, will result in a complete coverage from the IWA for charge=1 to the IWA for charge=3. 

\subsection{IWA for a FN with mirror symmetry}

The LBT FN is one variation of FNs with mirror symmetry, and we refer to it as a dual-aperture FN. The IWA of this type of FN is determined by the baseline of the interferometer. To make a connection between a dual-aperture FN and a dual-aperture VFN, we add a vortex plate in the optical system to investigate if the charge number would affect the IWA of a dual-aperture system. Equivalently, the dual-aperture FN corresponds to a charge=0 VFN. 

As shown in Fig. \ref{fig:single_dual_aperture_vfn_iwa} Right, the angular separation with the highest throughput does not move out as charge increases. An analogy to help understand the dependence is a traditional dual-aperture interferometer: the locations of the first null and/or the first constructive interferogram remain the same as long as the baseline remains the same, regardless of the diameter of the sub-aperture. The net effect of increasing the charge number in a dual-aperture system is to reduce the peak throughput without adding the benefit of relaxing the sensitivity to low-order aberrations. 


\subsection{Comparing LBT FN and GMT VFN}
\label{sec:lbt_gmt}
An LBT FN (B = 22.8 m) can provide similar spatial resolution to that of GMT (B = 25.2 m). Although the light collecting power of LBT is $(7/2)=3.5$ lower than GMT, a FN can have $\gtrapprox2$ higher planet throughput and lower sensitivity to low-order aberrations than a VFN. These factors would significantly reduce the exposure time for the LBT FN (Eq. \ref{eq:t_L}). In order to understand the trade-off between light collecting power, the planet throughput, and the sensitivity to low-order aberrations, we study two specific cases: an LBT FN and a GMT VFN. In the comparison, we assume everything is the same except for the light collecting area, planet throughput, and the sensitivity to low-order aberrations.

The sensitivity to lower-order aberrations for the GMT VFN is calculated the same way as described in \S \ref{sec:metrics} (see also Table \ref{tab:bi}) with one exception: wavefront errors are applied across a pupil that consists of 7 sub-apertures rather than one sub-aperture. The numerical results are in agreement with~\citep{Ruane2018}. 

\begin{figure}[h]
\plotone{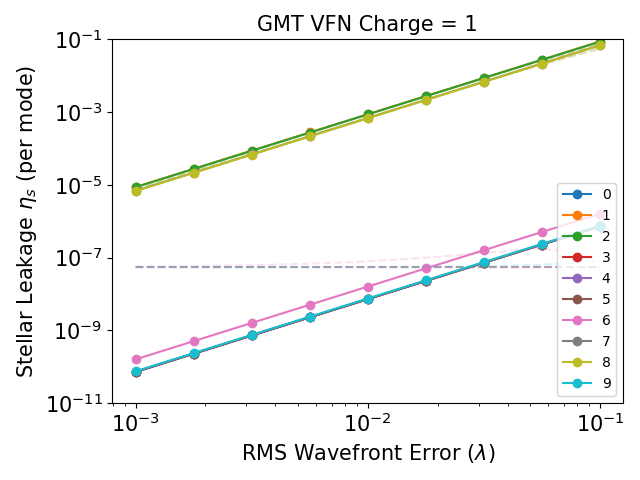}
\caption{Starlight suppression level vs. RMS wavefront error for each Zernike mode for a GMT VFN. Solid lines are fitting results, dashed lines connect data points in numerical simulations. Quantitative relationships between $\eta_s$ and wavefront error are given in Table \ref{tab:bi}. \label{fig:eta_s_abb_1_gmt}}
\end{figure}

Fig. \ref{fig:eta_s_abb_1_gmt} shows the dependence on RMS wavefront error per Zernike mode. Only Zernike modes with $l=\pm1$, i.e., with Zernike indices of 1, 2, 7, and 8 are coupled into a single-mode fiber. This is also confirmed in~\citet{Ruane2018, Ruane2019}. Given the non-zero numerical values for other Zernike modes that should not be theoretically decoupled from a single-mode fiber, we conclude that the uncertainties of $b_i$ coefficients that are reported in Table \S \ref{tab:bi} are $\sim$0.01. 

Using 55 Cnc c as an example, the total exposure time to reach a 5-$\sigma$ detection for a $K$-band LBT FN is {4.8 hours (Table \ref{tab:55cnc_c_k}) vs. 19.7 hour for a VFN at GMT (Table \ref{tab:55cnc_c_k_gmt}).} Despite a factor of $\sim$4 lower light collecting power, the LBT FN needs only a factor of $\sim$4 shorter exposure time to reach the same detection significance as a VFN on GMT. The loss in effective aperture size is out-weighted by the increase of planet throughput and the decrease in low-order aberrations (Eq. \ref{eq:t_L}). By comparing Table \ref{tab:55cnc_c_k} and Table \ref{tab:55cnc_c_k_gmt}, the difference in planet throughput is 27.0\% for the LBT FN vs. 15.8\% for the GMT VFN. A factor of 1.7 translates into a difference of 2.9 in exposure time based on Eq. \ref{eq:t_L}. In addition, the starlight suppression level for the LBT is 6.4 times better than that for the GMT VFN. This further reduces the LBT FN exposure time by a factor of 6.4 times. Together, these factors explain why the LBT FN outperforms the GMT VFN by a factor of 4. 

The above comparison of performance in $K$ holds as long as the total exposure time is dominated by the exposure time to overcome the low-order aberrations. However, this is no longer true in $L$ band, in which case the dominating noise source is the thermal background. Although the exposure time to overcome the thermal background $\tau_{\rm{bg}}$ is sensitive to the planet throughput, i.e., $\propto\eta_p^{-2}$,~\citep[see Eq. 12 in ][]{Ruane2018}, $\tau_{\rm{bg}}$ is also proportional to the solid angle subtended by the fiber, which is a factor of $\sim$3 smaller for the GMT case than the LBT case. Therefore, the $L$-band exposure time ratio between the LBT FN and the GMT VFN for 55 Cnc c is 2.3 (Table \ref{tab:55cnc_c_l} and Table \ref{tab:55cnc_c_l_gmt}). While the ratio is not as promising as the $K$-band case, it is nonetheless better than 3.5, which is from simply scaling the effective aperture size. 

\section{Summary}
\label{sec:summary}

We present a concept of combining nuling interferometry with single-mode fiber-fed high-resolution spectroscopy, i.e., the dual-aperture FN, which can be applied to current-generation 8-10 meter telescopes. The dual-aperture FN provides spatial resolution that is comparable to that of future ELTs, and therefore enables several unique science cases for 8-10 m telescopes before the era of ELTs in 2030s and future space interferometric missions. 

We conduct numerical simulations as a proof of concept in \S \ref{sec:sim}. We quantify planet throughput as a function of angular separation, single-mode fiber core size, and central obscuration due to a secondary mirror (Fig. \ref{fig:eta_p_sep}). We use a merit system, which is based on the required exposure time, to evaluate the performance of the dual-aperture FN in \S \ref{sec:metrics}. In particular, we quantify the sensitivity of starlight leakage to low-order aberrations as expressed by Zernike modes (Fig. \ref{fig:eta_s_abb} and Table \ref{tab:bi}). 

Because of the superior spatial resolution brought by interferometry and the planet sensitivity brought by single-mode fiber-fed high-resolution spectroscopy, a number of science cases are enabled by the dual-aperture FN, including (1) follow-up spectroscopic observations on exoplanet systems that are detected by the radial velocity technique (\S \ref{sec:rv_planet_ground}); (2) searching for planets in debris-disk system (\S \ref{sec:dusty}); and (3) direct spectroscopy for biosignatures in rocky planets around nearby M stars (\S \ref{sec:space_inter}). 

Targets for each case are given in Table \ref{tab:rv_followup}, Table \ref{tab:disk_sys}, and Table \ref{tab:space_rv}. Specific examples are discussed and general exposure time calculators are provided on GitHub\footnote{\url{https://github.com/wj198414/VFN}}. In all cases, we find that the dual-aperture FN is a viable pathway to achieve the science goals with reasonable telescope time investments. 

We compare FN and VFN in \S \ref{sec:fn_vfn}. The two concepts are connected by pupil-plane phase manipulation in order to achieve starlight suppression. We use GMT as an example to illustrate the connection and the synergy between the two concepts. We also compare the performance of a GMT VFN and an LBT FN. The comparison showcases that the LBT FN is indeed bridging the gaps between 8-10 telescopes and future ELTs because of its spatial resolution, planet throughput and sensitivity.  


\noindent{\it Acknowledgements} 
{We thank the anonymous referee whose comments and suggestions greatly improve the manuscript.} We would like to thank Elodie Choquet, Karl Stapelfeldt, and Bin Ren for helpful discusions on the debris-disk science case, Dan Echeverri, Garreth Ruane, and Dimitri Mawet for insights into VFN, Steve Ertel, Jordan Stone, and Amali Vaz for useful information on LBT and LBTI, and Bertrand Mennesson for discussing fiber nullers. 

\appendix

\section{Appendix information}

\subsection{55 Cnc c $K$-band}

\begin{deluxetable*}{lcll}
\tablecaption{55 Cnc c $K$-band simulation detail. \label{tab:55cnc_c_k}}
\tablewidth{0pt}
\tablehead{
}
\startdata
\multicolumn{4}{c}{\bf{Star Parameters}}\\
star name               &=& 55 Cnc c                      &\\
magnitude               &=& 4.01   &mag\\
distance                &=& 12.59      &pc\\
star radius             &=& 0.94   &solar radii\\
planet-star contrast    &=& 5.47e-07&\\
planet-star separation  &=& 19.17    &mas\\
\multicolumn{4}{c}{\bf{Telescope Parameters}}\\
aperture                &=& 8.40     &meter\\
baseline                &=& 22.40    &meter\\
wavelength              &=& 2.00     &um\\
lambda/D                &=& 18.42    &mas\\
\multicolumn{4}{c}{\bf{Coronagraph Parameters}}\\
filter                  &=& K         &\\
starlight suppression   &=& 4.81e-04&\\
planet throughput       &=& 2.70e-01&\\
bmn array               &=& [2.09 0.48 2.26 0.26 2.41 0.36 2.38 0.03 2.36 0.06]                                                 &\\
aberration array        &=& [0.01 0.01 0.01 0.01 0.01 0.01 0.01 0.01 0.01 0.01]                                                 &\\
gamma array     &=& [2 2 4 2 4 2 4 2 4 2]                                                                               &\\
\multicolumn{4}{c}{\bf{Spectrograph Parameters}}\\
required SNR            &=& 5.0     &\\
background              &=& 12.20  &mag / arcsec**2\\
spectral resolution     &=& 100000  &\\
pixel sampling rate     &=& 3.0   &pixel per resolution element\\
system throughput       &=& 0.100 &\\
fiber size              &=& 9.84e-04 &arcsec**2\\
\multicolumn{4}{c}{\bf{Detector Parameters}}\\
dark current            &=& 1.00e-02 &electron/pixel/s\\
readout noise           &=& 2.0000   &electron/read\\
quantum efficiency      &=& 0.950 &\\
well depth              &=& 65536    &ADU\\
\multicolumn{4}{c}{\bf{Exposure Times}}\\
t0                      &=& 4.08471e+09 &s\\
finite star time        &=& 6.24177e+05 &s\\
low order time          &=& 2.70424e+07 &s\\
background time         &=& 2.94258e+04 &s\\
readout noise time      &=& 1.65054e+03 &s\\
dark current time       &=& 1.64610e+05 &s\\
total time              &=& 2.78622e+07 &s\\
total time              &=& 7739.51    &hour\\
boost factor            &=& 1600.0   &\\
total time              &=& 4.84       &hour\\
\enddata
\end{deluxetable*}

\subsection{55 Cnc c $L$-band}

\begin{deluxetable*}{lcll}
\tablecaption{55 Cnc c $L$-band simulation detail. \label{tab:55cnc_c_l}}
\tablewidth{0pt}
\tablehead{
}
\startdata
\multicolumn{4}{c}{\bf{Star Parameters}}\\
star name               &=& 55 Cnc c                      &\\
magnitude               &=& 4.00   &mag\\
distance                &=& 12.59      &pc\\
star radius             &=& 0.94   &solar radii\\
planet-star contrast    &=& 5.47e-07&\\
planet-star separation  &=& 19.17    &mas\\
\multicolumn{4}{c}{\bf{Telescope Parameters}}\\
aperture                &=& 8.40     &meter\\
baseline                &=& 22.40    &meter\\
wavelength              &=& 3.50     &um\\
lambda/D                &=& 32.23    &mas\\
\multicolumn{4}{c}{\bf{Coronagraph Parameters}}\\
filter                  &=& L         &\\
starlight suppression   &=& 4.81e-04&\\
planet throughput       &=& 3.10e-01&\\
bmn array               &=& [2.09 0.48 2.26 0.26 2.41 0.36 2.38 0.03 2.36 0.06]                                                 &\\
aberration array        &=& [0.01 0.01 0.01 0.01 0.01 0.01 0.01 0.01 0.01 0.01]                                                 &\\
gamma array     &=& [2 2 4 2 4 2 4 2 4 2]                                                                               &\\
\multicolumn{4}{c}{\bf{Spectrograph Parameters}}\\
required SNR            &=& 5.0     &\\
background              &=& 2.00   &mag / arcsec**2\\
spectral resolution     &=& 100000  &\\
pixel sampling rate     &=& 3.0   &pixel per resolution element\\
system throughput       &=& 0.100 &\\
fiber size              &=& 3.02e-03 &arcsec**2\\
\multicolumn{4}{c}{\bf{Detector Parameters}}\\
dark current            &=& 1.00e-02 &electron/pixel/s\\
readout noise           &=& 2.0000   &electron/read\\
quantum efficiency      &=& 0.950 &\\
well depth              &=& 65536    &ADU\\
\multicolumn{4}{c}{\bf{Exposure Times}}\\
t0                      &=& 6.33963e+09 &s\\
finite star time        &=& 2.38512e+05 &s\\
low order time          &=& 3.16463e+07 &s\\
background time         &=& 1.25223e+09 &s\\
readout noise time      &=& 1.93154e+03 &s\\
dark current time       &=& 2.98976e+05 &s\\
total time              &=& 1.28441e+09 &s\\
total time              &=& 356780.86  &hour\\
boost factor            &=& 1225.0   &\\
total time              &=& 291.25     &hour\\
\enddata
\end{deluxetable*}

\subsection{HD 104860 debris system in $K$-band}

\begin{deluxetable*}{lcll}
\tablecaption{HD 104860 $K$-band simulation detail. \label{tab:hd104960_k}}
\tablewidth{0pt}
\tablehead{
}
\startdata
\multicolumn{4}{c}{\bf{Star Parameters}}\\
star name               &=& HD 104860                     &\\
magnitude               &=& 7.01   &mag\\
distance                &=& 47.90      &pc\\
star radius             &=& 1.00   &solar radii\\
planet-star contrast    &=& 1.00e-06&\\
\multicolumn{4}{c}{\bf{Telescope Parameters}}\\
aperture                &=& 8.40     &meter\\
baseline                &=& 22.40    &meter\\
wavelength              &=& 2.00     &um\\
lambda/D                &=& 18.42    &mas\\
\multicolumn{4}{c}{\bf{Coronagraph Parameters}}\\
filter                  &=& K         &\\
starlight suppression   &=& 4.81e-04&\\
planet throughput       &=& 2.00e-01&\\
bmn array               &=& [2.09 0.48 2.26 0.26 2.41 0.36 2.38 0.03 2.36 0.06]                                                 &\\
aberration array        &=& [0.01 0.01 0.01 0.01 0.01 0.01 0.01 0.01 0.01 0.01]                                                 &\\
gamma array     &=& [2 2 4 2 4 2 4 2 4 2]                                                                               &\\
\multicolumn{4}{c}{\bf{Spectrograph Parameters}}\\
required SNR            &=& 5.0     &\\
background              &=& 12.20  &mag / arcsec**2\\
spectral resolution     &=& 100000  &\\
pixel sampling rate     &=& 3.0   &pixel per resolution element\\
system throughput       &=& 0.100 &\\
fiber size              &=& 9.84e-04 &arcsec**2\\
\multicolumn{4}{c}{\bf{Detector Parameters}}\\
dark current            &=& 1.00e-02 &electron/pixel/s\\
readout noise           &=& 2.0000   &electron/read\\
quantum efficiency      &=& 0.950 &\\
well depth              &=& 65536    &ADU\\
\multicolumn{4}{c}{\bf{Exposure Times}}\\
t0                      &=& 1.92587e+10 &s\\
finite star time        &=& 4.18144e+05 &s\\
low order time          &=& 2.31706e+08 &s\\
background time         &=& 3.97814e+06 &s\\
readout noise time      &=& 1.41422e+04 &s\\
dark current time       &=& 2.22539e+07 &s\\
total time              &=& 2.58371e+08 &s\\
total time              &=& 71769.62   &hour\\
boost factor            &=& 1600.0   &\\
total time              &=& 44.86      &hour\\
\enddata
\end{deluxetable*}

\subsection{GJ 1061 b in $K$-band}

\begin{deluxetable*}{lcll}
\tablecaption{GJ 1061 b $K$-band simulation detail. \label{tab:gj1061b_k}}
\tablewidth{0pt}
\tablehead{
}
\startdata
\multicolumn{4}{c}{\bf{Star Parameters}}\\
star name               &=& GJ 1061 b                     &\\
magnitude               &=& 6.61   &mag\\
distance                &=& 3.67       &pc\\
star radius             &=& 0.16   &solar radii\\
planet-star contrast    &=& 1.47e-06&\\
planet-star separation  &=& 5.72     &mas\\
\multicolumn{4}{c}{\bf{Telescope Parameters}}\\
aperture                &=& 4.00     &meter\\
baseline                &=& 50.00    &meter\\
wavelength              &=& 2.00     &um\\
lambda/D                &=& 8.25     &mas\\
\multicolumn{4}{c}{\bf{Coronagraph Parameters}}\\
filter                  &=& K         &\\
starlight suppression   &=& 4.81e-04&\\
planet throughput       &=& 3.50e-01&\\
bmn array               &=& [2.09 0.48 2.26 0.26 2.41 0.36 2.38 0.03 2.36 0.06]                                                 &\\
aberration array        &=& [0.01 0.01 0.01 0.01 0.01 0.01 0.01 0.01 0.01 0.01]                                                 &\\
gamma array     &=& [2 2 4 2 4 2 4 2 4 2]                                                                               &\\
\multicolumn{4}{c}{\bf{Spectrograph Parameters}}\\
required SNR            &=& 5.0     &\\
background              &=& 20.40  &mag / arcsec**2\\
spectral resolution     &=& 100000  &\\
pixel sampling rate     &=& 3.0   &pixel per resolution element\\
system throughput       &=& 0.100 &\\
fiber size              &=& 9.26e-04 &arcsec**2\\
\multicolumn{4}{c}{\bf{Detector Parameters}}\\
dark current            &=& 1.00e-02 &electron/pixel/s\\
readout noise           &=& 2.0000   &electron/read\\
quantum efficiency      &=& 0.950 &\\
well depth              &=& 65536    &ADU\\
\multicolumn{4}{c}{\bf{Exposure Times}}\\
t0                      &=& 2.70121e+10 &s\\
finite star time        &=& 8.87692e+05 &s\\
low order time          &=& 1.06119e+08 &s\\
background time         &=& 6.22105e+02 &s\\
readout noise time      &=& 6.47699e+03 &s\\
dark current time       &=& 3.10952e+07 &s\\
total time              &=& 1.38109e+08 &s\\
total time              &=& 38363.59   &hour\\
boost factor            &=& 1600.0   &\\
total time              &=& 23.98      &hour\\
\enddata
\end{deluxetable*}

\subsection{Proxima Cen b in $L$-band}

\begin{deluxetable*}{lcll}
\tablecaption{Proxima Cen b $L$-band simulation detail. \label{tab:pro_cen_b_l}}
\tablewidth{0pt}
\tablehead{
}
\startdata
\multicolumn{4}{c}{\bf{Star Parameters}}\\
star name               &=& Proxima Cen b                 &\\
magnitude               &=& 4.38   &mag\\
distance                &=& 1.30       &pc\\
star radius             &=& 0.14   &solar radii\\
planet-star contrast    &=& 2.65e-07&\\
planet-star separation  &=& 37.31    &mas\\
\multicolumn{4}{c}{\bf{Telescope Parameters}}\\
aperture                &=& 4.00     &meter\\
baseline                &=& 20.00    &meter\\
wavelength              &=& 3.50     &um\\
lambda/D                &=& 36.10    &mas\\
\multicolumn{4}{c}{\bf{Coronagraph Parameters}}\\
filter                  &=& L         &\\
starlight suppression   &=& 4.81e-04&\\
planet throughput       &=& 3.50e-01&\\
bmn array               &=& [2.09 0.48 2.26 0.26 2.41 0.36 2.38 0.03 2.36 0.06]                                                 &\\
aberration array        &=& [0.01 0.01 0.01 0.01 0.01 0.01 0.01 0.01 0.01 0.01]                                                 &\\
gamma array     &=& [2 2 4 2 4 2 4 2 4 2]                                                                               &\\
\multicolumn{4}{c}{\bf{Spectrograph Parameters}}\\
required SNR            &=& 5.0     &\\
background              &=& 19.50  &mag / arcsec**2\\
spectral resolution     &=& 100000  &\\
pixel sampling rate     &=& 3.0   &pixel per resolution element\\
system throughput       &=& 0.100 &\\
fiber size              &=& 7.09e-03 &arcsec**2\\
\multicolumn{4}{c}{\bf{Detector Parameters}}\\
dark current            &=& 1.00e-02 &electron/pixel/s\\
readout noise           &=& 2.0000   &electron/read\\
quantum efficiency      &=& 0.950 &\\
well depth              &=& 65536    &ADU\\
\multicolumn{4}{c}{\bf{Exposure Times}}\\
t0                      &=& 1.69050e+11 &s\\
finite star time        &=& 4.42756e+06 &s\\
low order time          &=& 6.64125e+08 &s\\
background time         &=& 8.78841e+03 &s\\
readout noise time      &=& 4.05350e+04 &s\\
dark current time       &=& 3.93740e+07 &s\\
total time              &=& 7.07976e+08 &s\\
total time              &=& 196660.05  &hour\\
boost factor            &=& 1225.0   &\\
total time              &=& 160.54     &hour\\
\enddata
\end{deluxetable*}

\subsection{55 Cnc c $K$-band for GMT}

\begin{deluxetable*}{lcll}
\tablecaption{55 Cnc c $K$-band simulation detail. \label{tab:55cnc_c_k_gmt}}
\tablewidth{0pt}
\tablehead{
}
\startdata
\multicolumn{4}{c}{\bf{Star Parameters}}\\
star name               &=& 55 Cnc c                      &\\
magnitude               &=& 4.01   &mag\\
distance                &=& 12.59      &pc\\
star radius             &=& 0.94   &solar radii\\
planet-star contrast    &=& 5.47e-07&\\
planet-star separation  &=& 19.17    &mas\\
\multicolumn{4}{c}{\bf{Telescope Parameters}}\\
aperture                &=& 25.20    &meter\\
baseline                &=& 25.20    &meter\\
wavelength              &=& 2.00     &um\\
lambda/D                &=& 16.37    &mas\\
\multicolumn{4}{c}{\bf{Coronagraph Parameters}}\\
filter                  &=& K         &\\
starlight suppression   &=& 3.09e-03&\\
planet throughput       &=& 1.58e-01&\\
bmn array               &=& [0.01 2.94 2.94 0.01 0.01 0.01 0.01 2.61 2.6  0.01]                                                 &\\
aberration array        &=& [0.01 0.01 0.01 0.01 0.01 0.01 0.01 0.01 0.01 0.01]                                                 &\\
gamma array     &=& [2 2 2 2 2 2 2 2 2 2]                                                                               &\\
\multicolumn{4}{c}{\bf{Spectrograph Parameters}}\\
required SNR            &=& 5.0     &\\
background              &=& 12.20  &mag / arcsec**2\\
spectral resolution     &=& 100000  &\\
pixel sampling rate     &=& 3.0   &pixel per resolution element\\
system throughput       &=& 0.100 &\\
fiber size              &=& 4.13e-04 &arcsec**2\\
\multicolumn{4}{c}{\bf{Detector Parameters}}\\
dark current            &=& 1.00e-02 &electron/pixel/s\\
readout noise           &=& 2.0000   &electron/read\\
quantum efficiency      &=& 0.950 &\\
well depth              &=& 65536    &ADU\\
\multicolumn{4}{c}{\bf{Exposure Times}}\\
t0                      &=& 9.07713e+08 &s\\
finite star time        &=& 1.36229e+06 &s\\
low order time          &=& 1.12139e+08 &s\\
background time         &=& 7.97362e+03 &s\\
readout noise time      &=& 6.84440e+03 &s\\
dark current time       &=& 1.18277e+04 &s\\
total time              &=& 1.13528e+08 &s\\
total time              &=& 31535.45   &hour\\
boost factor            &=& 1600.0   &\\
total time              &=& 19.71      &hour\\
\enddata
\end{deluxetable*}

\subsection{55 Cnc c $L$-band for GMT}

\begin{deluxetable*}{lcll}
\tablecaption{55 Cnc c $L$-band simulation detail. \label{tab:55cnc_c_l_gmt}}
\tablewidth{0pt}
\tablehead{
}
\startdata
\multicolumn{4}{c}{\bf{Star Parameters}}\\
star name               &=& 55 Cnc c                      &\\
magnitude               &=& 4.00   &mag\\
distance                &=& 12.59      &pc\\
star radius             &=& 0.94   &solar radii\\
planet-star contrast    &=& 5.47e-07&\\
planet-star separation  &=& 19.17    &mas\\
\multicolumn{4}{c}{\bf{Telescope Parameters}}\\
aperture                &=& 25.20    &meter\\
baseline                &=& 25.20    &meter\\
wavelength              &=& 3.50     &um\\
lambda/D                &=& 28.65    &mas\\
\multicolumn{4}{c}{\bf{Coronagraph Parameters}}\\
filter                  &=& L         &\\
starlight suppression   &=& 3.09e-03&\\
planet throughput       &=& 1.67e-01&\\
bmn array               &=& [0.01 2.94 2.94 0.01 0.01 0.01 0.01 2.61 2.6  0.01]                                                 &\\
aberration array        &=& [0.01 0.01 0.01 0.01 0.01 0.01 0.01 0.01 0.01 0.01]                                                 &\\
gamma array     &=& [2 2 2 2 2 2 2 2 2 2]                                                                               &\\
\multicolumn{4}{c}{\bf{Spectrograph Parameters}}\\
required SNR            &=& 5.0     &\\
background              &=& 2.00   &mag / arcsec**2\\
spectral resolution     &=& 100000  &\\
pixel sampling rate     &=& 3.0   &pixel per resolution element\\
system throughput       &=& 0.100 &\\
fiber size              &=& 1.26e-03 &arcsec**2\\
\multicolumn{4}{c}{\bf{Detector Parameters}}\\
dark current            &=& 1.00e-02 &electron/pixel/s\\
readout noise           &=& 2.0000   &electron/read\\
quantum efficiency      &=& 0.950 &\\
well depth              &=& 65536    &ADU\\
\multicolumn{4}{c}{\bf{Exposure Times}}\\
t0                      &=& 1.40881e+09 &s\\
finite star time        &=& 6.16827e+05 &s\\
low order time          &=& 1.55499e+08 &s\\
background time         &=& 4.02071e+08 &s\\
readout noise time      &=& 9.49089e+03 &s\\
dark current time       &=& 2.54550e+04 &s\\
total time              &=& 5.58222e+08 &s\\
total time              &=& 155061.60  &hour\\
boost factor            &=& 1225.0   &\\
total time              &=& 126.58     &hour\\
\enddata
\end{deluxetable*}

\bibliography{sample63}{}
\bibliographystyle{aasjournal}



\end{CJK*}
\end{document}